# Echo disappears: momentum term structure and cyclic information in turnover

Haoyu Wang[1*], Junpeng Di[2], Yuegu Xie[2]

1. School of Management, Xiamen University, Xiamen, Fujian, China
2. Institute of Economy, Shanghai Academy of Social Science, Shanghai, China

**Corresponding author (*):** Haoyu Wang, Xiamen University, School of Management, 422 Siming South Road, Siming District, Xiamen, Fujian Province, China. Zip code: 361005. Email: why_ecofiliter@126.com. Phone: +86 18217137360.

## Abstract

We extract cyclic information in turnover and find it can explain the momentum "echo". The reversal in recent month momentum is the key factor that cancels out the recent month momentum and excluding it makes the "echo" regress to a damped shape. Both rational and behavioral theories can explain the reversal. This study is the first explanation of the momentum "echo" in U.S. stock markets.

## 1. Introduction

It seemingly becomes a widely accepted empirical cognition that momentum has an echo-like structure, indicating that the intermediate month return is dominant in price momentum. As Novy-Marx (2012) claims that firms' return from 12 to 7 months prior to portfolio formation primarily drives momentum but strategies based on recent past performance generate less profitable returns, this empirical finding yields theoretical research a confusing situation, because both rational theory and behavior theory predict a damped momentum and the echo-like structure is quite difficult to be theoretically reached. Several possible explanations, like the 12-month effect (Jegadeesh 1993; Heston and Sadka, 2008), earnings momentum (Chan et al., 1996), and disposition effects (Grinblatt and Han, 2005), fail to explain this phenomenon empirically (Novy-Marx, 2012). Consequently, this unreasonable inconsistency between empirical findings and theories has not been solved but is a central challenge to momentum research.

We find the "echo" is caused by a combination of reversal effects in recent month momentum. The reversals are agglomerated in high cyclic turnover stocks in short horizons (or scales) and cannot be detected if using overall turnover. Excluding the reversals in momentum, the momentum "echo" regresses to a damped shape. The reversals, spanned by Fama-French three factors plus Pastor and Stambaugh (2003) liquidity factor, have two potential explanations, a rational one and a behavioral one. The rational one attributes the reversals to hedging the market risk because the reversals have negative significant loadings on the market risk factor, for which Barinov (2014) has claimed that aggregate volatility risk embeds in turnover premiums. Behavior one is more plausible that the reversals are resilience from overconfidence (Statman et al., 2006; Chou et al., 2013). Overconfident investors having past good performance trade more but realize their biased self-attribution when the price returns to average and as the market performs better, their biased self-attribute becomes more significant and makes the resilience more rapid.

We use the wavelet method to extract the turnover information in horizons (scales). Financial markets are driven by information, whose functions have their certain horizons and flow together in a common pool, and distinguishing and extracting them at their scales are quite determinant. For example, prices and returns have been proven to have frequency-specific premiums (Neuhierl and Varneskov, 2021; Bandi et al., 2021). Since turnover and price are twins that impact each other (Lee and Swaminathan, 2000; Statman et al., 2006), it is natural that turnover may have frequency-specific information as well. Precisely, Statman et al. (2006), using vector autoregression analysis, have found a time structure between the price impact and turnover but their method is difficult to apply in cross-section analysis because they cannot distinguish price impacts and generate corresponding turnover responses. Even if they can do so, their method cannot distinguish between short-term and long-term impacts which have different mechanisms. Frequency-based methods, especially wavelet family methods, can overcome these shortcomings. Wavelet family methods can detect local and conditional information in time series which, compared with unconditional models, generates more precision in asset pricing, especially for information-driven context, because information-driven shocks have their certain horizons and occurrence time which are difficult to detect in overall time series when they mix each other. Bandi et al. (2021) have established a spectral asset pricing model based on the wavelet method and found heterogeneity between return frequencies.

The main contributions of this study are fruitful. First, it coordinates theories and empirical findings. No theory can explain the "echo" in momentum. Both rational theory and behavior theory predict momentum in a damped shape. Indeed, when we extract and exclude the reversals mixed in the recent month momentum, the momentum term structure regresses to the damped shape consistent with momentum theories. The reversals have two potential explanations, hedging market risk or resilience of overconfidence.

Second, it explores the cyclic information in turnover. Since the interaction between price and turnover has been widely documented and cyclic information in price plays a determinant role in asset pricing models, it is a natural extension that cyclic information in turnover is important in driving the asset return. Turnover is mostly recognized as a sentiment factor but is also considered as a proxy of other mechanisms like different opinions, volatility risk, and arbitrage risk. However, there exists a time structure among these mechanisms. Sentiment driven by overconfidence usually has a short time horizon while different opinions plus short-sale constraints tend to have a longer horizon. Turnover is an aggregate indicator of these mechanisms which can be extracted by the frequency-dependent method according to their characteristic scales.

Third, it refreshes some empirical cognitions on the momentum term structure. Goyal and Wahal (2015) have a creative guess that the short-term reversal included in the last two months distorts recent month momentum which is quite hard to be tenable from the perspective of our explanation because excluding the last two months in momentum also excludes the most significant short-term momentum since momentum has a damped shape and this would offset exclusion of the reversal.

## 2. Data and methodology

We use CRSP and COMPUSAT datasets including NYSE, AMEX, and NASDAQ stocks from 1969 to 2020, and to avoid illiquidity, stocks whose price is below 5 dollars in that month are excluded. Following Gao and Ritter (2010) to eliminate double counting, the NASDAQ turnover is divided by 2.0 prior to January 2001, by 1.8 for the rest of 2001, by 1.6 for 2002-2003, and left unchanged thereafter. Cyclic turnovers are decomposed from stock turnover time series using Daubechies 2 (DB2) wavelet. For robustness, we use the last three months average cyclic turnover. The market factor and Fama-French three factors are collected from Kenneth French's website. Pastor and Stambaugh liquidity factor is collected from Robert F. Stambaugh's personal

## 3. "Echo" of momentum term-structure

Novy-Marx (2012) claims intermediate month return is the primary factor driving momentum because intermediate month momentum has a much larger premium than recent month momentum. In his original article, premiums are calculated by Fama-Macbeth regression on CRSP and COMPUSAT datasets from 1927 to 2010. In this article, under the premise of basic precision, we use CRSP and COMPUSAT datasets from 1969 to 2020 and univariate portfolio analysis to reproduce his conclusion. All portfolio analyses in this article use one monthly holding period. Table 1 presents momentum premiums of recent month return and intermediate month return.

Table 1. Premiums of recent month return and intermediate month return. Univariate analysis allocates stocks into 10 groups according to sort variables. Returns in percentage are listed in the upper lines and weighted by market value. T-statistics are listed in subsequent lines and adjusted by the Newey-West method. $r_{6,2}$ is the recent month return and $r_{12,7}$ is the intermediate month return. $\alpha$ is the intercept controlled by Fama-French 3 factors model.

| Group | L | 2 | 3 | 4 | 5 | 6 | 7 | 8 | 9 | H | Diff |
|---|---|---|---|---|---|---|---|---|---|---|---|
| $r_{6,2}$ | 0.709 | 0.936 | 0.975 | 0.901 | 0.948 | 0.928 | 0.88 | 0.934 | 0.917 | 1.253 | 0.544 |
|  | 2.445 | 4.29 | 5.199 | 4.669 | 5.773 | 5.245 | 4.996 | 4.787 | 4.72 | 5.139 | 2.6 |
| $\alpha_{6,2}$ | -0.027 | 0.289 | 0.398 | 0.36 | 0.422 | 0.42 | 0.369 | 0.432 | 0.403 | 0.741 | 0.769 |
|  | -0.19 | 2.583 | 4.466 | 5.214 | 6.154 | 5.943 | 5.655 | 6.142 | 4.943 | 6.33 | 3.847 |
| $r_{12,7}$ | 0.5 | 0.719 | 0.84 | 0.843 | 0.935 | 0.886 | 1.012 | 1.087 | 1.208 | 1.402 | 0.902 |
|  | 1.884 | 3.19 | 4.276 | 4.622 | 5.405 | 5.248 | 5.835 | 6.244 | 5.701 | 5.394 | 4.219 |
| $\alpha_{12,7}$ | -0.225 | 0.08 | 0.236 | 0.268 | 0.39 | 0.365 | 0.512 | 0.579 | 0.702 | 0.875 | 1.1 |
|  | -2.001 | 0.814 | 2.839 | 4.085 | 5.729 | 5.199 | 6.952 | 7.689 | 7.578 | 6.959 | 5.919 |

Table 1 shows that the recent month return produces a 0.5% premium and the intermediate month return produces a 0.9% premium. The difference in alphas between the two returns remains the same. Values of premiums are consistent with Novy-Marx's calculation which guarantees our results are unbiased though different methods are applied.

## 4. Cyclic information in turnover and momentum term-structure

Cyclic information in turnover probably has a great influence on the momentum term structure. First, we extract cyclic information in turnover at specific orthogonal scales. We treat turnover as a combination of information with different scales and use the DB2 wavelet method to extract their behavior at certain scales. The wavelet method is a popular and general method to separate a time series into its corresponding frequency behavior which is called scale. Through scales of decomposed turnover, we can analyze different components of turnover. In this article, we decompose the turnover series into 7 scales. Scales are orthogonal indicating that their cyclic characteristics are independent as well. Each scale has a specific cycle and the corresponding decomposed series depicts the behavior of turnover in that cycle. The correspondence of scales and cycles by wavelet decomposition is in table 2.

Table 2. Correspondence of scales and cycles by wavelet decomposition. Scales are decomposed using the DB2 wavelet method and cycles are multiple of two. For example, scale 6 depicts turnover characteristics in 0~2 months cycle. Scale 5 depicts turnover characteristics in 2~4 months cycle.

| Scale | Cycle |
|---|---|
| 0 | >64months |
| 1 | 32~64months |
| 2 | 16~32months |
| 3 | 8~16months |
| 4 | 4~8months |
| 5 | 2~4months |
| 6 | 0~2months |

Having extracted components of cyclic turnover, we use cyclic turnovers as sort variables. For robustness, we use the last three months average cyclic turnovers as sort variables for each cycle, and cyclic turnovers, if not especially mentioned, refer to average cyclic turnovers in this article. We calculate their covariance coefficients with recent and intermediate month returns to ensure enough stocks in every group in bivariate portfolio analysis. The covariance coefficients are reported in table 3.

Table 3. Covariance coefficients of variables. 'Turn_AVE_x' is the last three months average cyclic turnover at scale x. '$r_{m,n}$' is the aggregate return of the last $m$ months to last $n$ months.

|  | Turn_AVE_0 | Turn_AVE_1 | Turn_AVE_2 | Turn_AVE_3 | Turn_AVE_4 | Turn_AVE_5 | Turn_AVE_6 | $r_{6,2}$ | $r_{12,7}$ |
|---|---|---|---|---|---|---|---|---|---|
| Turn_AVE_0 | 1.000 | -0.064 | -0.047 | -0.029 | -0.015 | -0.010 | -0.003 | 0.024 | 0.032 |
| Turn_AVE_1 | -0.064 | 1.000 | 0.007 | 0.003 | -0.001 | -0.001 | 0.000 | 0.032 | 0.051 |
| Turn_AVE_2 | -0.047 | 0.007 | 1.000 | 0.005 | 0.006 | 0.003 | 0.001 | 0.068 | 0.100 |
| Turn_AVE_3 | -0.029 | 0.003 | 0.005 | 1.000 | 0.012 | 0.019 | 0.006 | 0.105 | 0.057 |
| Turn_AVE_4 | -0.015 | -0.001 | 0.006 | 0.012 | 1.000 | 0.044 | 0.034 | 0.114 | -0.046 |
| Turn_AVE_5 | -0.010 | -0.001 | 0.003 | 0.019 | 0.044 | 1.000 | 0.069 | 0.056 | -0.009 |
| Turn_AVE_6 | -0.003 | 0.000 | 0.001 | 0.006 | 0.034 | 0.069 | 1.000 | 0.010 | 0.002 |
| $r_{6,2}$ | 0.024 | 0.032 | 0.068 | 0.105 | 0.114 | 0.056 | 0.010 | 1.000 | 0.013 |
| $r_{12,7}$ | 0.032 | 0.051 | 0.100 | 0.057 | -0.046 | -0.009 | 0.002 | 0.013 | 1.000 |

In table 3, the coefficients of cyclic turnovers are close to zero which validates their orthogonality. The orthogonality excludes the interaction of cyclic turnovers at different scales which simplifies our analysis. The coefficients between cyclic turnovers and the term-structure return are close to zero which ensures enough stocks in every single portfolio.

Having introduced the essential information, we conduct a bivariate sort analysis. The column variables are always recent month return and intermediate month return. The row variables are cyclic turnovers from scale 0 to scale 6 and each sort consumes one scale of cyclic turnovers. For brevity in table 4, we only present the key differences between the two momentum terms which are among scale 4 and scale 5 and the complete results are in appendices.

Table 4. Bivariate sort portfolio analysis. In panel A, the row sort variable is the cyclic turnover at scale 4 and the column sort variable is the recent month return (RR) and intermediate month return (IR). In panel B, the row sort variable is the cyclic turnover at scale 5 and the column sort variable is the same as in panel A. The portfolio returns in percentage are on the left-hand side and the t-statistics are on the right-hand side. "Diff" is the high-minus-low portfolio return.

|  | Estimate (percent per month) | | | | | | Test Statistics | | | | | |
|---|---|---|---|---|---|---|---|---|---|---|---|---|
| Panel A | | | | | | Scale 4 | | | | | | |
| | | | | | | Recent month momentum | | | | | | |
| | RR1 | RR2 | RR3 | RR4 | RR5 | Diff | RR1 | RR2 | RR3 | RR4 | RR5 | Diff |
| T1 | 0.667 | 0.799 | 0.73 | 1.009 | 1.759 | 1.093 | 1.868 | 3.257 | 3.235 | 4.088 | 6.094 | 3.716 |
| T2 | 0.467 | 0.64 | 0.985 | 0.757 | 1.258 | 0.791 | 1.578 | 3.083 | 5 | 3.463 | 5.206 | 3.371 |
| T3 | 0.475 | 0.72 | 0.919 | 0.998 | 1.104 | 0.63 | 1.67 | 3.855 | 5.98 | 5.171 | 5.464 | 2.796 |
| T4 | 0.66 | 0.786 | 0.815 | 0.883 | 1.262 | 0.602 | 2.56 | 3.812 | 4.604 | 4.998 | 5.353 | 2.429 |
| T5 | 0.903 | 1.065 | 1.008 | 0.95 | 1.14 | 0.237 | 3.661 | 5.632 | 5.727 | 5.188 | 5.051 | 1.264 |
| T6 | 0.83 | 0.998 | 1.035 | 1.045 | 1.096 | 0.266 | 3.341 | 4.921 | 5.708 | 5.351 | 4.581 | 1.304 |
| T7 | 1.089 | 1.026 | 1.084 | 0.841 | 0.955 | -0.134 | 4.517 | 5.466 | 5.851 | 4.013 | 3.995 | -0.717 |
| T8 | 0.906 | 1.105 | 1.156 | 0.963 | 1.094 | 0.188 | 3.439 | 5.265 | 6.268 | 4.946 | 4.899 | 0.963 |
| T9 | 1.045 | 1.159 | 0.977 | 0.874 | 0.993 | -0.052 | 4.194 | 5.392 | 4.61 | 4.586 | 4.171 | -0.28 |
| T10 | 1.215 | 1.269 | 1.128 | 0.952 | 0.748 | -0.467 | 4.275 | 6.131 | 5.888 | 4.753 | 3.061 | -2.291 |
| Diff | 0.548 | 0.47 | 0.398 | -0.057 | -1.011 | -1.56 | 2.598 | 2.359 | 2.57 | -0.31 | -4.554 | -5.976 |
| | | | | | | Intermediate month momentum | | | | | | |
| | IR1 | IR2 | IR3 | IR4 | IR5 | Diff | IR1 | IR2 | IR3 | IR4 | IR5 | Diff |
| T1 | 0.687 | 0.636 | 0.674 | 1.063 | 1.525 | 0.839 | 2.342 | 2.39 | 3.109 | 4.193 | 5.802 | 3.848 |
| T2 | 0.384 | 0.575 | 0.816 | 1.044 | 1.117 | 0.733 | 1.423 | 2.645 | 4.078 | 5.894 | 4.435 | 3.441 |
| T3 | 0.638 | 0.792 | 0.739 | 0.882 | 1.166 | 0.529 | 2.727 | 4.057 | 3.859 | 4.694 | 5.486 | 2.666 |
| T4 | 0.564 | 0.755 | 0.863 | 1.025 | 1.452 | 0.888 | 2.377 | 3.839 | 4.73 | 5.376 | 6.451 | 4.592 |
| T5 | 0.791 | 0.833 | 1.099 | 1.252 | 1.506 | 0.715 | 3.156 | 4.486 | 6.339 | 7.42 | 5.988 | 3.292 |
| T6 | 0.613 | 0.9 | 1.022 | 1.118 | 1.204 | 0.591 | 2.498 | 4.477 | 5.521 | 6.222 | 4.499 | 2.496 |
| T7 | 0.616 | 1.078 | 0.918 | 1.004 | 1.289 | 0.674 | 2.296 | 5.524 | 5.201 | 5.281 | 5.351 | 3.078 |
| T8 | 0.742 | 1.003 | 0.972 | 1.088 | 1.352 | 0.611 | 2.758 | 5.318 | 4.795 | 5.992 | 5.538 | 2.72 |

| | | | | | | | | | | | |
|---|---|---|---|---|---|---|---|---|---|---|---|
| T9 | 0.793 | 0.823 | 0.926 | 0.953 | 1.211 | 0.417 | 3.232 | 3.887 | 4.627 | 4.325 | 4.905 | 2.214 |
| T10 | 0.886 | 1.094 | 1.055 | 1.216 | 1.069 | 0.183 | 3.095 | 4.78 | 5.31 | 5.847 | 3.791 | 0.764 |
| Diff | 0.199 | 0.458 | 0.38 | 0.153 | -0.456 | -0.655 | 1.059 | 2.267 | 2.228 | 0.948 | -2.334 | -2.677 |

| Panel B | Scale 5 | | | | | | | | | | |
|---|---|---|---|---|---|---|---|---|---|---|---|
| | Recent month momentum | | | | | | | | | | |
| | RR1 | RR2 | RR3 | RR4 | RR5 | Diff | RR1 | RR2 | RR3 | RR4 | RR5 | Diff |
| T1 | 0.984 | 0.976 | 1.007 | 1.179 | 1.769 | 0.785 | 2.525 | 3.473 | 3.515 | 4.911 | 5.352 | 2.66 |
| T2 | 0.503 | 0.701 | 0.821 | 0.882 | 1.387 | 0.883 | 1.799 | 2.991 | 3.687 | 3.772 | 5.278 | 4.333 |
| T3 | 0.658 | 0.699 | 1.004 | 0.971 | 1.217 | 0.559 | 2.381 | 3.476 | 5.037 | 4.885 | 4.58 | 2.309 |
| T4 | 0.578 | 0.974 | 0.955 | 0.944 | 0.989 | 0.412 | 2.093 | 4.805 | 5.028 | 4.906 | 4.172 | 1.985 |
| T5 | 0.881 | 0.883 | 0.947 | 0.917 | 1.119 | 0.238 | 3.487 | 4.091 | 5.561 | 4.678 | 5.002 | 1.056 |
| T6 | 0.968 | 0.924 | 0.913 | 0.879 | 1.171 | 0.203 | 4.078 | 4.344 | 4.863 | 3.962 | 5.183 | 1.064 |
| T7 | 0.922 | 1.041 | 1.166 | 1.101 | 0.989 | 0.067 | 3.479 | 5.169 | 7.682 | 5.665 | 4.398 | 0.299 |
| T8 | 0.896 | 1.085 | 0.954 | 0.771 | 0.935 | 0.039 | 3.597 | 5.79 | 5.434 | 4.182 | 4.121 | 0.211 |
| T9 | 1.079 | 1.341 | 1 | 0.791 | 1.026 | -0.053 | 4.222 | 7.384 | 5.824 | 4.377 | 5.013 | -0.271 |
| T10 | 1.09 | 1.118 | 1.026 | 0.774 | 0.629 | -0.461 | 4.287 | 6.25 | 5.964 | 4.215 | 3.019 | -2.414 |
| Diff | 0.105 | 0.142 | 0.02 | -0.406 | -1.141 | -1.246 | 0.374 | 0.627 | 0.081 | -1.982 | -4.524 | -4.707 |
| | Intermediate month momentum | | | | | | | | | | |
| | IR1 | IR2 | IR3 | IR4 | IR5 | Diff | IR1 | IR2 | IR3 | IR4 | IR5 | Diff |
| T1 | 0.963 | 0.916 | 1.027 | 1.223 | 1.648 | 0.686 | 2.749 | 2.914 | 3.85 | 4.527 | 5.009 | 2.408 |
| T2 | 0.713 | 0.748 | 0.797 | 0.86 | 1.095 | 0.382 | 2.61 | 3.155 | 3.612 | 3.79 | 3.932 | 1.386 |
| T3 | 0.796 | 0.895 | 0.72 | 1.021 | 1.174 | 0.378 | 2.847 | 4.277 | 3.642 | 5.66 | 4.435 | 1.508 |
| T4 | 0.64 | 0.802 | 0.893 | 0.951 | 1.262 | 0.621 | 2.456 | 4.097 | 4.647 | 4.779 | 5.09 | 3.131 |
| T5 | 0.663 | 0.863 | 0.843 | 1.099 | 1.224 | 0.562 | 2.743 | 4.989 | 4.313 | 5.576 | 4.66 | 2.407 |
| T6 | 0.726 | 0.948 | 1.084 | 1.043 | 1.356 | 0.631 | 2.806 | 4.706 | 6.332 | 5.379 | 5.448 | 2.871 |
| T7 | 0.66 | 0.852 | 1.135 | 1.183 | 1.361 | 0.701 | 2.825 | 4.395 | 6.928 | 6.404 | 5.75 | 3.821 |
| T8 | 0.446 | 0.773 | 0.925 | 1.059 | 1.325 | 0.879 | 1.702 | 3.855 | 5.402 | 5.717 | 6.056 | 4.107 |
| T9 | 0.524 | 0.965 | 0.998 | 1.022 | 1.452 | 0.928 | 2.187 | 5.546 | 5.362 | 5.496 | 6.936 | 4.757 |
| T10 | 0.476 | 0.767 | 0.946 | 1.036 | 1.226 | 0.751 | 2.015 | 3.908 | 5.481 | 6.102 | 5.435 | 4.16 |
| Diff | -0.487 | -0.148 | -0.08 | -0.187 | -0.422 | 0.065 | -1.797 | -0.572 | -0.356 | -0.808 | -1.679 | 0.226 |

Recent month momentum at scales 4 and 5 mostly concentrates on small turnover stocks. At scale 4, high-minus-low portfolios, T1-Diff, T2-Diff, T3-Diff, and T4-Diff, have 1.09%, 0.79%, 0.63% and 0.60% premiums. At scale 5, high-minus-low portfolios, T1-Diff, T2-Diff, and T3-Diff, have 0.79%, 0.88%, and 0.56% premiums. Reversal of recent month momentum at short scales mostly concentrates on large turnover stocks. High-minus-low portfolios, T10-Diff at scale 4 and T10-Diff at scale 5, have -0.47% and -0.46% premiums. It is an interesting finding that reversal hides in momentum and what drives this phenomenon is discussed in section 7. Intermediate month momentum at scale 4 concentrates on small cyclic turnover stocks. On scale 5, the momentum has no significant pattern.

The comparison of cyclic turnover's effect between recent month momentum and intermediate month momentum is evident to explain the "echo" shape of momentum. At scales 4 and 5, the cyclic turnover's effect is plausibly determinant that large cyclic turnover stocks have a reversal in recent month return but have momentum in intermediate month return while small cyclic turnover stocks have more premiums in recent month return but have fewer premiums in intermediate month return. Therefore, it is plausible that the reversals in short cycles mediate recent month momentum whose distortion generates the "echo" shape of the momentum term structure.

## 5. Spanning test

Having examined the influence of cyclic turnovers on momentum term structure, we make a spanning test to examine its effect on the momentum term structure. In table 5, we use the portfolio return of reversals (T10-Diff of recent month momentum at scale 4 and scale 5) and control groups (T10-Diff of intermediate month momentum at scale 3 and scale 4) in table 4 as independent variables and portfolio return of recent month momentum and intermediate month momentum as dependent variables to check the performance of term-structure after controlled by reversals.

In table 1, recent month momentum $MOM_{6,2}$ and intermediate month momentum $MOM_{12,7}$ have premiums 0.54% and 0.92% and alphas 0.77% and 1.10%, respectively. In regression (6) of table 5, after controlled by reversals of cycle 4 and cycle 5 and FF3, the recent month momentum has a premium of 1.03% which is close to the intermediate month momentum's alpha and is larger than its premium (0.77%) controlled by control groups return. Stepwise regression results are similar. These results validate reversals of cyclic turnovers are dominants of the "echo" shape of the momentum term structure. Indeed, after being controlled by the reversals, the "echo" regresses to a damped shape.

Table 5. Performance of momentum term-structure controlled by reversal. Panel A reports the regression of recent month momentum, $MOM_{6,2}$, on reversals in cycle4, "reversal-cycle4", and in cycle 5, "reversal-cycle5". T-statistics in parentheses and $adj.R^2$ are reported. Panel B reports the regression of intermediate month momentum, $MOM_{12,7}$, on control groups return in cycle3, "control group-cycle3", and in cycle 4, "control group-cycle4". T-statistics in parentheses and $adj.R^2$ are reported. "FF3 control" is Fama-French three factors model.

| Panel A | | | | | | |
|---|---|---|---|---|---|---|
| | | | $y = MOM_{6,2}$ | | | |
| Independent variable | (1) | (2) | (3) | (4) | (5) | (6) |
| Intercept | 0.91 | 0.97 | 0.94 | 1 | 0.99 | 1.03 |
| | (6.636) | (7.055) | (6.517) | (6.923) | (8.256) | (8.515) |
| Reversal-Cycle4 | 0.7867 | 0.7759 | | | 0.4943 | 0.4907 |
| | (32.864) | (31.743) | | | (16.862) | (16.711) |
| Reversal-Cycle5 | | | 0.8558 | 0.8442 | 0.467 | 0.4659 |
| | | | (30.443) | (29.413) | (14.214) | (14.142) |
| FF3 control | No | Yes | No | Yes | No | Yes |
| $Adj.R^2$ | 0.627 | 0.632 | 0.591 | 0.597 | 0.716 | 0.719 |
| Panel B | | | | | | |
| | | | $y = MOM_{12,7}$ | | | |
| Independent variable | (1) | (2) | (3) | (4) | (5) | (6) |
| Intercept | 0.8 | 0.95 | 0.69 | 0.78 | 0.68 | 0.77 |
| | (4.954) | (5.984) | (4.705) | (5.29) | (5.037) | (5.6644) |
| ControlGroup-Cycle4 | 0.5542 | 0.5267 | | | 0.2918 | 0.2941 |
| | (18.59) | (17.876) | | | (9.789) | (9.908) |
| ControlGroup-Cycle3 | | | 0.6175 | 0.5864 | 0.4661 | 0.4364 |
| | | | (23.934) | (22.044) | (16.287) | (15.029) |
| FF3 control | No | Yes | No | Yes | No | Yes |
| $Adj.R^2$ | 0.35 | 0.393 | 0.471 | 0.483 | 0.54 | 0.551 |

## 6. Fama-Macbeth regression

In Novy-Marx 's (2012) original article, he uses Fama-Macbeth regression to estimate risk premiums of the momentum term structure. In table 6, We repeat his method using our dataset. We use cyclic turnovers and original turnovers to control their influence on the term structure and keep other control variables identical to those in Novy-Marx (2012). Novy-Marx (2012) uses the last month return, log size, and log book-to-market ratio as control variables. It is worth mentioning that our dataset is from 1969 to 2020 which is different from his 1927~2010 dataset and it is not surprising that our risk-premiums are different but the aforementioned robust portfolio returns are the same as his estimates. For robust estimation, we use two types of Fama-Macbeth regression one contains an intercept in regression and the other does not. In model (3), recent month return and intermediate month return have premiums, 0.49% (t-statistics=2.20) and 0.57% (t-statistics=3.12), indicating the "echo" shape of the term structure. After being controlled by turnover in the model (2), the two premiums become 0.53% (2.71) and 0.49% (2.77). After controlled by cyclic turnovers of cycle 3~cycle 5, the two premiums invert to 0.73% (3.21) and 0.53% (2.96) indicating the "echo" shape transforms to a damped shape. Models (4)~(6) including the regression intercept have the same results that the two premiums transform from 0.42% (1.98) and 0.39% (2.24) to 0.64% (3.01) and 0.35% (2.07). Therefore, we validate the deterministic role of cyclic turnovers in the momentum term structure.

Table 6. Fama-Macbeth regression. '$Turnover_{cyclex}$' is cyclic turnover at scale $x$. Model (1)~(3) have the same regression model as that of Novy-Marx (2012). Model (4)~(6) add an intercept in the original model. Model (1) and (4) control cyclic turnovers. Model (2) and (5) control original turnover. Other control variables are the same as those of Novy-Marx (2012). Other parameters are the same as previous tables.

| | $r_{tj} = \beta' x_{tj} + \varepsilon_{tj}$ | | | $r_{tj} = \alpha + \beta' x_{tj} + \varepsilon_{tj}$ | | |
|---|---|---|---|---|---|---|
| Independent variable | (1) | (2) | (3) | (4) | (5) | (6) |
| RR | 0.73 | 0.53 | 0.49 | 0.64 | 0.51 | 0.42 |
| | 3.215 | 2.709 | 2.206 | 3.011 | 2.609 | 1.981 |
| IR | 0.53 | 0.49 | 0.57 | 0.35 | 0.35 | 0.39 |
| | 2.963 | 2.766 | 3.121 | 2.07 | 2.017 | 2.24 |
| $Turnover_{all}$ | | 0.43 | | | 0.22 | |
| | | 4.292 | | | 2.37 | |
| $Turnover_{cycle_3}$ | -0.2 | | | -0.1 | | |
| | -1.372 | | | -0.731 | | |
| $Turnover_{cycle_4}$ | -1.48 | | | -1.32 | | |
| | -10.073 | | | -9.163 | | |
| $Turnover_{cycle_5}$ | -5.95 | | | -5.72 | | |
| | -14.406 | | | -14.583 | | |
| $r_{1,0}$ | -3.24 | -3.6 | -3.25 | -3.54 | -3.68 | -3.56 |
| | -8.045 | -8.911 | -8.111 | -8.594 | -9.031 | -8.666 |
| $\log(ME)$ | 0.14 | 0.01 | 0.14 | -0.41 | -0.47 | -0.41 |
| | 8.136 | 0.282 | 7.984 | -10.408 | -9.399 | -10.082 |
| $\log(BM)$ | 0.43 | 0.47 | 0.43 | 0.15 | 0.17 | 0.14 |
| | 4.578 | 5.046 | 4.436 | 1.636 | 2.062 | 1.502 |
| $Adj.R^2$ | 0.161 | 0.165 | 0.154 | 0.048 | 0.051 | 0.04 |
| $Average\ n$ | 2849 | 2849 | 2849 | 2849 | 2849 | 2849 |

## 7. Attribute analysis

We analyze the origin of the reversals. Whether the reversals originate from short-term reversal and whether it is rational or behavioral remains a discussion. Goyal and Wahal (2015) suppose it may be a continuation of short-term reversal though they cannot prove their conjecture. Our study provides a plausible explanation that they, excluding the last two months return, exclude reversal and momentum simultaneously and make it difficult to prove their conjecture. In subsequent analysis, we find the reversals have loadings on the market risk factor and both rational and behavioral explanations are plausible.

### 7.1. Short-term reversal

We test whether cyclic reversals come from short-term reversals. We use returns of reversal portfolios, T10-Diff of recent month momentum at scale 4 and 5 in table 4, as dependent variables and use market factor and short-term reversal factor as independent variables. Results in table 7 show cyclic turnovers cannot be explained by short-term reversal. Model (1) and (3) report significant negative intercepts, -0.4% (t-statistics=-1.92), -0.41% (t-statistics=-2.16), indicating controlling short-term reversals cannot eliminate premiums of cyclic reversals. Model (2) and (4) report intercepts, -0.32% (t-statistics=-1.49), -0.35% (t-statistics=-1.96), indicating adding market risk factor significantly improve interpretations but model (4) still has significant premiums. Therefore, cyclic reversals though have loadings on the short-term reversals but are not a shadow of it.

Table 7. Cyclic reversal and short-term reversal.

| | $Reversal - Cycle\ 4$ | | $Reversal - Cycle\ 5$ | |
|---|---|---|---|---|
| Independent variable | (1) | (2) | (3) | (4) |
| Intercept | -0.4* | -0.32 | -0.41** | -0.35* |
| | -1.921 | -1.496 | -2.163 | -1.969 |
| MKT | | -0.178*** | | -0.128 |
| | | -2.359 | | -1.64 |
| Reversal | 0.5798*** | 0.4969*** | 0.4243*** | 0.3646** |
| | 3.8 | 2.63 | 3.288 | 2.112 |

|  | $Adj.R^2$ | 0.055 | 0.073 | 0.036 | 0.047 |
|---|---|---|---|---|---|

## 7.2. CAPM and Fama-French three factors model

We test cyclic reversals' loadings on CAPM and Fama-French three factors model. We use returns of reversal portfolios, T10-Diff of recent month momentum at scale 4 and 5 in table 4, as dependent variables and use market factor, Fama-French three factors (FF3), and Pastor and Stambaugh liquidity factor as independent variables.

Results in table 8 show cyclic turnovers can be explained by FF3 plus liquidity. Model (1) and (4) report significant negative intercepts, -0.34% (t-statistics=-1.66) and -0.37% (t-statistics=-2.07), indicating controlling the market factor cannot eliminate premiums of cyclic reversals. Model (2) and (5) report intercepts, -0.27% (t-statistics=-1.29) and -0.28% (t-statistics=-1.51), indicating adding SMB and HML significantly improve interpretations. Model (3) and (6) report intercepts, -0.03% (t-statistics=-0.12) and -0.02% (t-statistics=0.12), indicating adding liquidity factor significantly improves interpretations and cyclic reversals no longer have significant premiums.

Therefore, after being controlled by FF3 plus liquidity factors, cyclic reversals have no significant premiums and have major loadings on the market factor. Since Barinov (2014) has claimed that aggregate volatility risk embeds in turnover premiums, the reversals can be hedging for the market risk factor. Alternatively, the reversals can be resilience from overconfidence (Statman et al., 2006). Overconfident investors having past good performance trade more but realize their biased self-attribution when the price returns to average. As their mispricing fades and the market performs better, their bias self-attribute becomes more significant and makes the resilience more rapid.

Table 8. Spanning test of cyclic reversal on CAPM, Fama-French three factors, and Pastor and Stambaugh liquidity factor.

|  | $Reversal - Cycle\ 4$ | | | $Reversal - Cycle\ 5$ | | |
|---|---|---|---|---|---|---|
| Independent variable | (1) | (2) | (3) | (4) | (5) | (6) |
| Intercept | -0.34* | -0.27 | -0.03 | -0.37** | -0.28 | 0.02 |
|  | -1.661 | -1.292 | -0.127 | -2.076 | -1.515 | 0.125 |
| MKT | -0.24*** | -0.29*** | -0.33*** | -0.17** | -0.24*** | -0.27*** |
|  | -3.162 | -4.7 | -4.81 | -2.346 | -3.454 | -3.721 |
| SMB |  | 0.12 | 0.11 |  | 0.15 | 0.14 |
|  |  | 0.727 | 0.648 |  | 0.789 | 0.75 |
| HML |  | -0.2 | -0.19 |  | -0.25* | -0.27* |
|  |  | -1.2 | -1.134 |  | -1.713 | -1.824 |
| Liquidity |  |  | 0.075 |  |  | 0.100** |
|  |  |  | 1.488 |  |  | 2.278 |
| $Adj.R^2$ | 0.035 | 0.048 | 0.054 | 0.022 | 0.05 | 0.067 |

## 8. Conclusion

We extract the cyclic information in turnover to explain the "echo" structure of momentum.

The reversal in recent month momentum is the main factor driving the "echo". After being controlled by the reversal, the momentum structure regresses to a damped shape. Our empirical results refresh the previous cognition of the momentum term structure and coordinate the inconsistency between momentum theories and empirical findings.